\documentclass{aa}
\usepackage{graphicx}
\newcommand{\teff}{T$_{\rm eff}$}
\newcommand{\vt}{$\xi$}
\begin{document}
\title{Detailed chemical composition of the open cluster IC 4651:  
The iron peak, $\alpha$ elements, and Li. 
\thanks{Based on observations collected at the ESO VLT, Paranal Observatory,
Chile}}

\subtitle{}
\author{L. Pasquini\inst{1}, S. Randich\inst{2}, M. Zoccali\inst{1}, V. 
Hill\inst{3}, C. Charbonnel\inst{4,5} and B. Nordstr\"om\inst{6,7}}
\offprints{lpasquin@eso.org}

\institute{European Southern Observatory, D-85748 Garching bei M\"unchen, 
Germany
\and
INAF--Osservatorio di Arcetri, Largo Enrico Fermi 5, I-50125 Firenze, Italy 
\and
GEPI (URA8111), Observatoire de Paris-Meudon, F-92195 Meudon, France 
\and
Observatoire de Gen{\`e}ve, CH-1290 Sauverny, Switzerland 
\and
LATT, OMP, CNRS UMR 5572, 14, av.E.Belin, 31400 Toulouse, France
\and 
    Niels Bohr Institute for Astronomy, Physics \& Geophysics, 
    Juliane Maries Vej 30, DK-2100 Copenhagen, Denmark
\and Lund Observatory, Box 43, S-221 00 Lund, Sweden }

\abstract{
We present a detailed chemical analysis of 22 stars along the colour-magnitude
sequence of the intermediate-age (1.7 Gyr) open cluster IC 4651, based on
high-resolution, high S/N ratio spectra from UVES/VLT. IC 4651 thus
becomes one of the few open clusters for which a detailed composition
analysis exists for stars spanning 3.5 magnitudes, from solar-type
main-sequence stars to giants above the RGB clump.  In a strict comparison 
with the Sun, we find for the cluster a well-defined Fe abundance 
of [Fe/H]= $0.10\pm$0.03 (internal errors), with a reddening {\it E(b-y)}= 
0.091.  We also derive abundances for 
the $\alpha$ elements Mg, Si, Ti, and Ca and find a moderate enhancement of 
the three former elements, in excellent agreement with the results for 
field stars of similar Fe abundance. Among the Fe group elements, Cr and Ni 
are slightly overabundant ([X/Fe]$\sim$0.05). 
As also observed in M67, the Na abundance among the giants is more 
than 0.2 dex higher than in the dwarfs. We interpret this effect as real, 
and  due to dredge-up of $^{23}$Na in the giants. 
\\
Four turnoff stars, all fairly rapid rotators ($v \sin i \leq$ 25 km s$^{-1}$), 
appear to have higher [Fe/H] than the other stars; our tests show that, while 
a spurious enhancement of 0.1 dex can be produced by the effect of high 
rotation on our spectral analysis, this cannot explain the whole difference. 
These stars appear overabundant because we tend to overestimate their 
effective temperatures by forcing excitation equilibrium.  \\
Li abundances have been computed for all the stars and show a well-defined 
pattern: the Solar-type stars have an almost constant Li abundance, just 
below that of the Hyades, and the Li-dip is pronounced and well determined.
Turnoff stars just above the dip have a `cosmic' Li abundance, but within 
a very small range of magnitudes (0.25 mag) higher on the turnoff, 
the Li level drops by more than a factor 10. 
This cannot be due just to dilution; rather some extra mixing is required. Among 
the giants, two probable clump stars show detectable Li, while all the 
other (likely) RGB stars do not - as is also observed in the similar cluster 
NGC3680. None of these patterns can be explained by classical stellar evolution 
models. Again, some extra mixing is required.
We show that rotating stellar models including the most recent 
developments for meridional circulation and turbulence by shear instabilities 
explain very well the behaviour of the lithium abundance along the 
colour-magnitude diagram of IC4651, including subgiant and giant phases. 
The possibility remains open that the giants exhibiting the highest lithium 
abundances are actual RGB bump stars which have just been through the so-called
``lithium flash".
}
\date {Received XXX; accepted XXX}
\authorrunning{L. Pasquini et al.}
\titlerunning{Abundances in IC 4651}
\maketitle
%
\section{Introduction}

Open clusters are key tracers of the chemical evolution of the Galaxy
and, at the same time, benchmarks for stellar evolution models. Despite 
the many advances in the detailed modelling of stellar evolution over the 
last few years (see e.g. Girardi et al. 2000), very few open clusters 
have been studied in a comprehensive manner. Membership studies have become 
available for several clusters, but the number of detailed chemical analyses 
remains very small (e.g., Gratton \cite{gratt00} and references therein).

The determination of heavy-element abundances and abundance ratios and 
their variation with time and Galactocentric distance is crucial for our 
understanding of the formation and chemical evolution of the Galactic disk 
(e.g., Edvardsson et al. 1993; Tosi 2000). Abundances in clusters also
provide a fundamental tool to constrain the relative importance of
Type I and Type II supernovae in the enrichment of the interstellar
medium and thus to gain insight in the star formation history in the
disk.  On the other hand, the abundance evolution of the
fragile elements Li and Be along the colour magnitude diagram (CMD) of a
cluster represents a precious testbench for stellar evolutionary theories, 
as the initial stellar mass is the only variable parameter within a cluster 
(see e.g. Pasquini et al. 2000; Deliyannis et al. 2000).

For these reasons, we decided to carry out a detailed abundance analysis of 
the rich, well-studied, intermediate age cluster IC~4651, based on high-quality 
spectra for 23 members obtained with the VLT/UVES. The Str{\"o}mgren photometry 
by Meibom (2000) and radial-velocity survey by Meibom et al. (2002) allowed 
us to select target stars known to be members, single, and with known rotational
velocity. It also provided the basic properties of the cluster (distance, 
reddening, overall metallicity, and age). 

In the present paper, we briefly summarize the properties of the cluster and 
describe the observations, data reduction and analysis. We then discuss the 
metal and $\alpha$ element abundances, while the last section before the 
conclusions is devoted to a detailed analysis of the evolution of lithium. 
The abundances of O and Be and the ${\mathrm C^{12}/C^{13}}$ isotopic ratio 
will be the subject of a future paper.

\section{Basic properties of IC~4651}
The early studies of IC~4651 by Eggen (1971), Lindoff (1972), and 
Anthony-Twarog et al. (1987, 1988, 2000) were superseded by the extensive 
$uvby$ photometry of 17,640 stars by Meibom (2000), supplemented by the 
radial-velocity survey of 104 stars by Meibom et al (2002). Notably, the 
wide field observed by Meibom (2000) roughly doubled the area and stellar 
content of the cluster relative to earlier studies; 
because the radial-velocity survey was initiated about a decade earlier, 
it is complete for the inner cluster only (to $V\sim14.5$). 

Meibom et al. (2002) found IC~4651 to have a reddening of 
{\it E(B-V)} = 0.10, a distance of 1 kpc ($(m-M)_0 = 10.03\pm0.1$), a metal 
abundance of [Fe/H] = +0.1 dex, and an age of $1.7\pm0.15$ Gyr; the mass of 
the cluster stars at the turnoff (TO) is $1.8 M_{\odot}$. The present 
stellar population of IC~4651 was estimated to be $\sim 650$ stars with a 
total mass of $\sim630 M_{\odot}$. IC~4651 was, however, also found to show 
moderate mass segregation. A comparison of the present and plausible initial 
mass functions indicated that it initially contained at least $\sim 8300$ 
stars with a total mass of $\sim 5300 M_{\odot}$. Of this mass, $\sim35\%$ 
has been lost during the evolution of the initially most massive stars, 
while the remaining $\sim53\%$ ($\sim93\%$ of the original low-mass stars) 
have evaporated from (the central part of) the cluster.

>From its space motion, Meibom et al. (2002) also computed the Galactic orbit 
of IC~4651. They found a fairly eccentric orbit ($e$ = 0.19), with a mean 
galactocentric orbital radius of 8.6 kpc and peri- and apogalactic distances 
7.0 and 10.2 kpc, respectively. IC~4651 is currently 7.1 kpc from the Galactic 
centre and 140 pc below the Galactic plane; its maximum distance from the 
plane is 190 pc.

The conclusion of Meibom et al. (2002) that IC~4651 is metal rich ([Fe/H] =
+0.1 dex) was based on the same type of $uvby\beta$ photometry as used by 
Nordstr\"om et al. (1997) to derive a similarly high metallicity for the 
roughly coeval cluster NGC~3680. However, for the latter cluster, Pasquini 
et al. (2001) found [Fe/H] = -0.17 from high-resolution spectroscopy, a 
result that appears to be corroborated by the recent photometry of 
Anthony-Twarog \& Twarog (2004). Thus, one goal of the present study is to 
check the iron abundance of IC~4651 by careful high-resolution spectroscopy.

\section{Sample selection and observations}

20 programme stars were selected for observation from the unpublished thesis 
by Meibom, pending the publication of the detailed results by Meibom (2000) 
and Meibom et al. (2002). Stars were chosen to be single and among the slower 
rotators in each magnitude bin, since an accurate analysis of very broad-lined 
spectra is not possible, in particular in the Be region. The stars were also 
chosen to sample a large section of the CMD; based on our previous experience 
with the twin cluster NGC3680 (Pasquini et al. 2001), it was clear that the 
region around the so called Li-dip (Boesgaard \& Tripicco 1986) at the turnoff 
(TO) would be particularly interesting; thus we took care to sample this region 
more densely than was possible in the more sparsely populated NGC~3680 
(Nordstr\"om et al. 1997). To these 20 stars we added the three stars previously 
analyzed by Randich et al. (2002), which were also observed with UVES in 
dichroic mode. 

\begin{figure}
\begin{center}
\includegraphics[width=9.0cm]{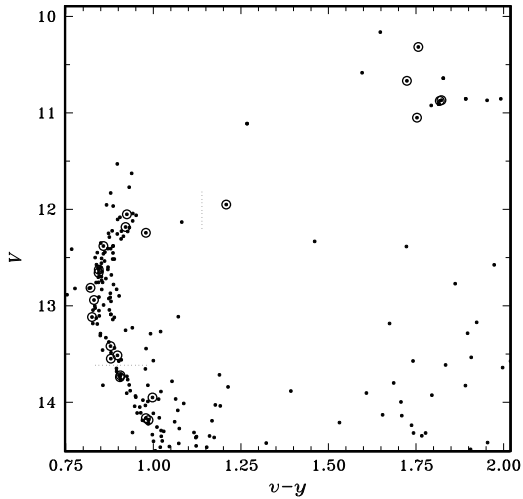}
\caption{Colour-magnitude diagram of IC 4651 from Meibom et al. (2002); our 
programme stars are indicated by circles. 
}
\label{CMD}
\end{center}
\end{figure}

Our sample (including the stars from Randich et al. 2002) is listed in Table 
1. Numbers less than 100 and with prefix ``E'' are from Eggen (1971); higher 
numbers with prefix ``T'' are from Anthony-Twarog et al. (1988). Str\"omgren 
photometry and rotational velocities are from Meibom (2000) and Meibom et al. 
(2002), while the {\it B-V} colors are from Anthony-Twarog et al. (1988); the 
$v$ magnitude for E98 by Meibom (2000) has been corrected for an apparent 
typographical error. The CMD of IC~4651 is shown in Fig. \ref{CMD}, with 
our sample stars circled for identification.

\begin{table}
\caption{Our sample stars in the giant, turnoff, and lower main-sequence 
regions. The Str\"omgren photometry is from Meibom (2000), {\it B-V} from Twarog 
et al. (1988), and v$\sin i$ (km~s$^{-1}$) from Meibom et al. (2002).}
\begin{tabular}{lcccccrl}
\hline
\hline
Name & $y$ & {\it b-y}  & $m_1$ & $c_1$ & {\it B-V} & v$\sin i$ \\
\hline   
E12   &  10.324 & 0.661 & 0.465 & 0.314 & 1.126 &  1.1  \\ 
E8    &  10.667 & 0.666 & 0.403 & 0.299 & 1.100 &  0.1  \\
E60   &  10.870 & 0.673 & 0.491 & 0.312 & 1.141 &  1.9  \\
E98   &  10.880 & 0.694 & 0.496 & 0.331 & 1.187 &  1.0  \\
T812  &  11.059 & 0.648 & 0.413 & 0.344 & 1.140 &  0.5  \\
E95   &  11.953 & 0.514 & 0.185 & 0.386 & 0.805 & 12.0  \\ \hline
E3    &  12.051 & 0.374 & 0.169 & 0.527 & 0.612 & 29.9  \\
E56   &  12.184 & 0.378 & 0.159 & 0.499 & 0.599 & 27.8  \\
E19   &  12.240 & 0.412 & 0.157 & 0.484 & 0.651 & 32.1  \\
E99   &  12.377 & 0.347 & 0.171 & 0.531 & 0.546 & 28.1  \\
E17   &  12.625 & 0.337 & 0.158 & 0.543 & 0.539 & 10.0: \\
E25   &  12.654 & 0.365 & 0.118 & 0.552 & 0.537 & 21.3  \\
E5    &  12.817 & 0.329 & 0.160 & 0.532 & 0.507 & 23.9  \\
T1228 &  12.941 & 0.344 & 0.130 & 0.540 & 0.536 &  5.7  \\
E14   &  13.119 & 0.335 & 0.154 & 0.507 & 0.516 & 34.1  \\
E34   &  13.418 & 0.358 & 0.156 & 0.459 & 0.566 & 25.2  \\
E15   &  13.515 & 0.374 & 0.146 & 0.447 & 0.563 & 10.0  \\
E79   &  13.549 & 0.360 & 0.154 & 0.444 & 0.514 & 21.8  \\ \hline
E64   &  13.722 & 0.370 & 0.160 & 0.430 & 0.594 &  9.2  \\
E86   &  13.739 & 0.383 & 0.143 & 0.431 & 0.626 & 14.0  \\
T2105 &  13.948 & 0.429 & 0.145 & 0.374 & 0.618 & 11.3  \\
E7    &  14.165 & 0.402 & 0.179 & 0.368 & 0.643 &  2.1  \\
E45   &  14.184 & 0.430 & 0.127 & 0.399 & 0.654 &  4.2  \\
\hline\hline
\end{tabular}
\end{table}

The observations were carried out at the ESO VLT Kueyen telescope and its 
UVES spectrograph (Dekker et al. 2000). We used UVES in dichroic mode, 
which allowed us to simultaneously observe the near-UV/blue and red spectral 
ranges. In order to obtain a good signal in the region of the UV Be lines 
(313 nm), much longer exposure times are required than at longer wavelengths. 
We thus chose to obtain several red exposures, using both CD\#3 centered
at 580~nm and CD\#4 centered at 860~nm, simultaneously with the UV ones; in 
this way we were able to observe the whole range 480-1000 nm in all the stars, 
with a gap in spectral coverage from 390 to 480~nm and very small gaps around 
580 and 860~nm.

The three stars previously analyzed by Randich et al. (2002), which were also 
observed with UVES in dichroic mode, had several red exposures each in the setup
centered around 580 nm. Each of these exposures was used to compute independent 
abundances, as if they pertained to different objects. This gives a good 
estimate of the internal random errors.

The observations were carried out in service mode between May and June 2001. 
Whereas in the UV a resolving power $R \sim45,000$ was used (1~arcsec slit), 
we opted for a very high resolution ($R=100,000$) in the red, using a 0.4 
arcsec wide slit. The typical S/N ratio of our spectra is 70--120 per pixel.
The data of Randich et al. were characterized by a similar S/N ratio,
but a lower resolution (R$\sim$45,000).

Both raw data and spectra reduced using the ESO--UVES pipeline (Ballester 
et al. 2000) were sent to us; after independently re-reducing several sample 
spectra we concluded that a complete, independent re-reduction was not 
warranted.

\section{Data analysis: Stellar parameters and determination of abundances}

\subsection{Stellar parameters and iron abundances }

Initial effective temperatures (\teff) were derived  
following the Alonso et al. (1996 and 1999) scales, based on {\it (b-y)} 
colours and assuming a reddening of {\it {\it E(b-y)}}= 0.072 (Meibom et al. 
2002). Using a TO mass of 1.8~M$_{\odot}$ and a distance modulus of $(m-M)_0$= 
10.03 (Meibom et al. 2002), we then computed surface gravities ($\log~g$) from 
mass, luminosity and \teff. For each star, \teff\ and the microturbulence 
velocity \vt\ have then been varied so as to eliminate any dependence of the 
Fe~{\sc i} abundance on the line excitation potential and equivalent width. 
Surface gravities were not varied by more than 0.3~dex.
Due to the good quality of the spectra and the many lines present in our 
large spectral range, we could avoid using lines in the flat portion of  
the curve of growth, i.e., equivalent widths larger than 100~m\AA. 
As mentioned, we did not attempt to vary log $g$ in order to have 
$\log~n$(Fe~{\sc i})$= \log~n$(Fe~{\sc ii}), but just tried to obtain
an ionization balance as good as possible while keeping a gravity
compatible with the position of the star in the CMD.  

Our analysis was carried out differentially with respect to the Sun
starting from measured equivalent widths, hence most of the systematic
effects should cancel out.  We shall see, however, that for some
elements this may not be completely true and some differences may be
present between dwarfs and giants.

Equivalent widths were measured by using an automatic line width
programme: DAOSPEC (Stetson et al. 2004) and typically 80--100 Fe~{\sc i} lines 
were used. The abundance calculations were performed with an updated and improved 
version of the original code described in Spite (1967). Edvardsson et al. (1993) 
model atmospheres were used for the dwarfs, while Gustafsson et al. (1975) 
models were employed for the giants.  For the subgiant E95 we
derived the Fe abundance using both models and found excellent
agreement (within 0.01 dex in Fe/H); thus, we do not expect any major
spurious effect by using the two sets of model atmospheres in the two
different evolutionary regimes.  Finally, for most elements we used the 
line list of Pancino et al. (2002); atomic parameters were mainly retrieved 
from the National Institute of Science and Technology Atomic Spectra Database.  
For Ca and Al, the data were taken instead from the VALD database (Kupka 
et al. \cite{vald2}).

\subsection{Iron abundances}

In Table~2 we list the initial photometric \teff, the final
(spectroscopic) adopted \teff, microturbulence \vt, and $\log~g$ values,
together with the derived Fe~{\sc i} and Fe~{\sc ii} abundances.  The first 
line of the table reports the solar abundances that we have determined by 
analyzing the UVES solar spectrum (www.eso.org/instruments/UVES) in the 
same fashion as for the IC~4651 stars. For the Sun we used an Edvardsson et al.  
model with \teff = 5750 K, $\log~g$= 4.4 and $\xi$= 0.9 km s$^{-1}$.  We note 
that our solar Fe~{\sc i} and Fe~{\sc ii} abundances are slightly higher than 
the canonical value ($\log~n$(Fe)= 7.52, Anders \& Grevesse \cite{ag89}). In 
particular, it is worth noting that while in the Sun the ionization equilibrium 
is obtained, this is not the case for most stars in IC~4651; for these Fe~{\sc 
ii} is usually lower than Fe~{\sc i} because we allowed the spectroscopic $\log 
g$ to depart no more than 0.3 dex from the photometric value.  

\begin{table}
\caption{Spectroscopic and  photometric \teff\ and [Fe/H] for  the Sun and the 
stars in IC~4651. For the Sun, we give $\log\epsilon$(Fe), while all cluster 
star abundances are relative to this value ([Fe/H]). Values in parenthesis were 
omitted when computing mean abundances. }
\begin{tabular}{lcccccr}
\hline
\hline
Star & T$_{\rm phot}$ & T$_{\rm spec}$ & $\log g$ & $\xi$ & [Fe~{\sc i}] & 
[Fe~{\sc ii}] \\ 
& (K) & (K) & & (km s$^{-1)}$ & & \\   
\hline 
Sun   &  5800 & 5750 & 4.4  &  0.9 &  7.57 &   7.59    \\
\hline 
E12   &  4940 & 5000 & 2.7  &  1.5 &  0.09 &  $-$0.13  \\
E8    &  4870 & 4900 & 2.7  &  1.3 &  0.09 &  $-$0.13  \\
E60   &  4850 & 4900 & 2.9  &  1.4 &  0.07 &  0.09     \\
E98   &  4870 & 4900 & 3.0  &  1.4 &  0.15 &  0.05     \\
E812  &  4830 & 5000 & 3.0  &  1.6 &  0.11 &  $-$0.05  \\
E95   &  5560 & 5800 & 3.5  &  1.7 &  0.07 &  $-$0.04  \\
Average &     &      &      &      & 0.097 &  $-0.035$ \\
$\sigma$ &    &      &      &      &  0.03 &  0.09     \\
\hline 
E3    &  6390 & 6550 & 3.9  &  2.1 &  0.27 &  0.16    \\
E56   &  6360 & 6500 & 3.9  &  2.1 &  0.11 &  $-$0.06 \\
E19   &  6127 & 6500 & 3.9  &  2.1 &  0.14 &  0.00    \\
E99   &  6656 & 6800 & 4.0  &  2.0 & (0.32)& (0.23)   \\
E17   &  6653 &      &      &      &       &          \\
E25   &  6514 & 6900 & 4.1  &  1.9 &  0.11 &  0.06    \\
E5    &  6750 & 7100 & 4.3  &  1.9 & (0.40)& (0.22)   \\
T1228 &  6600 & 6800 & 4.2  &  1.7 &  0.26 &  0.19    \\
E14  &  6690 & 6800 & 4.3  &  1.9  &  0.20 &  0.14    \\E34  &  6480 & 6850 & 
4.3  &  
1.9  & (0.45)& (0.45)   \\
E15  &  6380 & 6850 & 4.4  &  1.8  &  0.21 &  0.07    \\
E79  &  6460 & 6650 & 4.3  &  1.7  &  0.17 &  0.14    \\
Average &    &      &      &       &  0.18 &  0.09    \\
$\sigma$&    &      &      &       &  0.06 &  0.08    \\
\hline
*E64   &  6390 & 6650 & 4.3  &  1.7 &  0.09 &  $-$0.05 \\
*E86   &  6300 & 6600 & 4.3  &  1.7 &  0.13 &  0.06    \\
*T2105 &  5990 & 6350 & 4.3  &  1.1 &  0.04 &  $-$0.05 \\
*T2105 &  5990 & 6300 & 4.3  &  1.1 &  0.07 &  0.02    \\
*T2105 &  5990 & 6400 & 4.4  &  1.1 &  0.13 &  0.05    \\
*E7    &  6150 & 6300 & 4.3  &  1.1 &  0.13 &  0.12    \\
*E7    &  6150 & 6300 & 4.3  &  1.1 &  0.12 &  0.07    \\
*E45   &  5960 & 6350 & 4.3  &  1.1 &  0.12 &  0.08    \\
*E45   &  5960 & 6350 & 4.3  &  1.1 &  0.13 & 0.03     \\
Average &      &      &      &      & 0.107 & 0.04     \\
$\sigma$ &     &      &      &      &  0.03 & 0.06     \\
\hline 
\end{tabular}
\end{table}

The typical 1 $\sigma$ dispersion around the mean Fe I and Fe II abundances is 
0.1 dex, very uniform between slow-rotating dwarfs and giants. \\
Table 2 shows that: 
\begin{enumerate}
\item The abundances obtained from different spectra of the same star
(7 last entries in the table) are in very good agreement, 
within 0.09 dex for Fe~{\sc i} and 0.1~dex for Fe~{\sc ii} (peak to peak);

\item In most cases spectroscopic temperatures are higher than
photometric ones. While the difference between \teff(phot)
and \teff(spec) is within $\sim 150$~K for the giants, it can
be as high as 300--400~K for the other stars. Small discrepancies
could easily be explained, for instance by the characteristics of
the stellar model atmospheres adopted, but the differences
in T$_{\rm eff}$ are too high for most dwarfs to be explained
in this way;

\item The bulk of the stars (19, including all the giants) have Fe~{\sc i}
ranging between [Fe/H]= 0.03 and [Fe/H]= 0.26, while the remaining three
objects, all located at the TO, give more than 2 $\sigma$ higher
Fe~{\sc i} abundances;

\item The average Fe~{\sc i} abundances of giants and dwarfs are in
excellent agreement, while the average abundance of stars at the TO is
somewhat higher, even when not taking into account the three stars
mentioned above. More specifically, the six giants have a mean
[Fe~{\sc i}/H]= 0.10, to be compared with [Fe~{\sc i}/H]= 0.11 for the 
solar--type stars.  This is a very encouraging result,
confirming that not only the spectroscopic effective temperatures are
consistent, but also that the agreement between the two sets of model
atmospheres is very good.

However, eight stars, which are typically rather fast-rotating TO stars,
give a slightly larger metallicity ([Fe~{\sc i}/H]= 0.18), while the three
apparently most metal-rich stars (all hot and very fast rotators) give 
[Fe~{\sc i}/H] above 0.3 (E5, E34, E99). Anticipating the conclusions of our
discussion, we note that the abundances for all rapid rotators suffer 
substantial uncertainties, regardless whether their abundances agree with 
the other stars of the cluster. In addition, for TO stars our method is not 
sensitive enough to constrain \teff\ properly, and we shall see that, when 
adopting photometric \teff\ with proper reddening, we recover Fe abundances 
in very good agreement with the solar-type stars and the giants. 

\end{enumerate}

In the following, we discuss: {\it a)} possible reasons for
the overabundance measured among rapidly rotating stars at the TO; and
{\it b)} possible reasons for the difference between spectroscopic
and photometric \teff\ values.

\subsubsection{Hot fast-rotating turnoff stars}
Rapid rotation is a concern for abundance analyses based on
equivalent widths, since, when the stars rotate fast, the selected
lines may blend with nearby lines resulting in systematically higher
equivalent widths.  In addition, the measured scatter in [Fe/H] for fast 
rotators ($\sigma \sim$0.3 dex) is about 3 times higher than for the best 
(slow rotating) stars in IC4651. This large dispersion does not seriously 
degrade the formal uncertainties of the derived Fe abundance, because more 
than 60 Fe I lines were used in the determination.
Rather, the problem is that by excluding many lines in the analysis we
are no longer performing a strict solar comparison. In addition,
the solution becomes quite insensitive to variations in \teff, and the 
solution space becomes flat instead of showing a pronounced minimum.

We have confirmed this by artificially broadening the spectra of
three stars (the Sun, E45 and E25) and repeating the analysis.
We derive a spurously large microturbulence and an Fe abundance 
up to 0.1~dex higher than from the original spectra.  Both these points
are consistent with the data in Table 2, which shows that indeed for
virtually all TO stars the microturbulence velocity is higher than the 
canonical values, and that the (typically fast-rotating) TO stars have
[Fe/H]= 0.18, i.e. about 0.1 dex higher than the mean value derived
from the more slowly-rotating dwarfs and giants.  However, the
systematic effects due to higher rotation do not explain the large
differences between spectroscopic and photometric temperatures.

Finally, a synthetic and rotationally broadened spectrum for star E34 
around the Li~{\sc i} 6708~\AA~doublet was computed using MOOG (Sneden
\cite{sne}) and Kurucz (\cite{kuru}) model atmospheres.  The observed 
spectrum of E34 is shown in Fig. 2, together with two different 
synthetic spectra. The figure clearly shows that we cannot discriminate 
between the two solutions, which differ by 600 K and 0.3 dex in [Fe/H].

\begin{figure*}
\begin{center}
\includegraphics[width=12.0cm,height=12.0cm,angle=270]{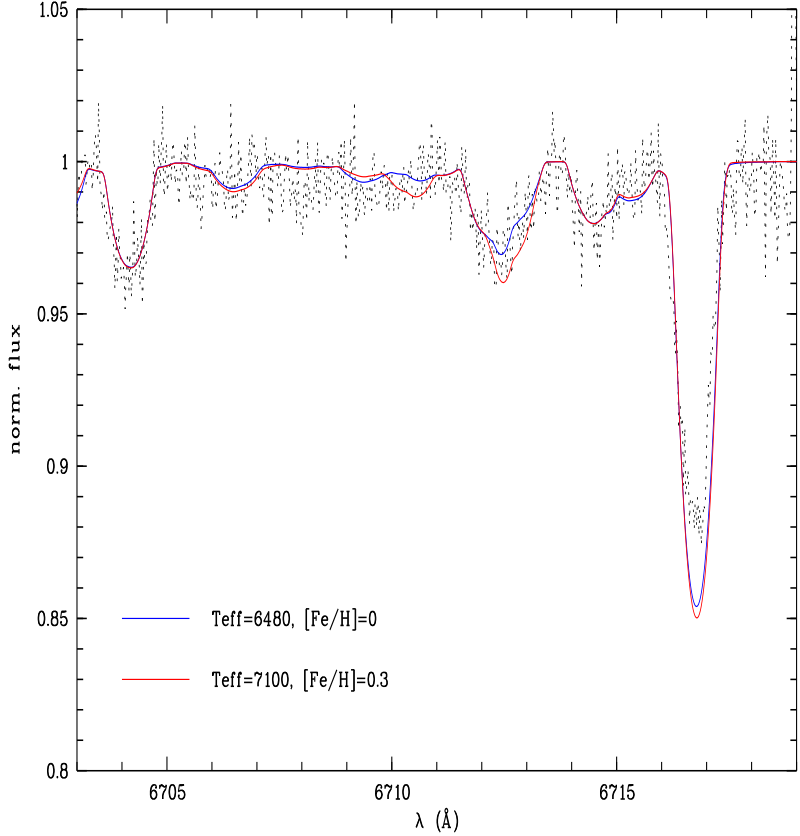}
\end{center}
\caption{ 
Two synthetic spectra for the Li region in the fast-rotating E34, differing by 
$\sim$ 600 K in \teff\ and 0.3 in [Fe/H]; the observed spectrum cannot 
discriminate between these two solutions. }
\label{synt}
\end{figure*}

Because of this uncertainty in the rapidly-rotating hot TO stars, we 
repeated the analysis using the photometric \teff\ derived with our best 
reddening estimate (cf. next section). The results are given in Fig. 3, where 
the Fe abundances of the 11 TO stars have been redetermined with the 
new photometric temperatures. The scatter in the Fe abundances is quite large 
(0.1 dex), but the mean ([Fe~I/H]= 0.15) is now only slightly higher and in very 
good agreement with [Fe/H]= 0.10 found for the giants and solar-type stars. 

\begin{table}
\caption{Photometric \teff\ and associated [Fe/H] values for the rapidly 
rotating TO stars in IC~4651, computed with our revised reddening estimate ({\it 
E(b-y)}=0.091). }
\begin{tabular}{lccccr}
\hline \hline
Star & T$_{\rm phot}$  & $\log g$ & $\xi$ & [Fe~{\sc i}] & [Fe~{\sc ii}]  \\   
\hline
E3   &     6550 &  3.9 &  2.1   &  0.27 &  0.16 \\ 
E56   &    6520 &  3.9 & 2.1    &  0.11 & -0.06 \\  
E19    &   6280 & 3.9  & 2.1    &  0.02 & 0.02  \\   
E99  &     6830 & 4.0  & 2.0    & 0.33  & 0.23  \\
E25   &    6680  & 4.0 & 2.0    & 0.00  & -0.02 \\
E5     &   6930  & 4.2 & 2.0    & 0.28  & 0.15  \\ 
T1228 &    6770  & 4.2 & 1.7    & 0.26  & 0.19   \\
E14    &   6860  & 4.2 & 1.9    & 0.23  & 0.18  \\
E34     &  6640  & 4.2 & 1.9    & 0.39  & 0.33  \\
E15   &    6540  & 4.2 & 1.7    & 0.08  & 0.06  \\ 
E79    &   6620  & 4.3 & 1.7    & 0.17  & 0.14  \\ \hline
Average &        &     &        &  0.15 & 0.13     \\
$\sigma$ &       &     &        &   0.1 & 0.1     \\
\hline ~
\end{tabular}
\end{table}

\subsubsection{Discrepancy between photometric and spectroscopic temperatures}

The average difference between spectroscopic and photometric \teff\ is 
290~and 100~K for the slow-rotating dwarfs and the giants, respectively. 
There are three possible explanations for the difference: {\it i)} the use
of an incorrect color--\teff\ calibration; {\it ii)} systematic
effects due to the adopted analysis code and model atmospheres; and
{\it iii)} an error in the estimate of reddening.

\noindent {\it i)} Using a different colour system and/or calibration, such 
as {\it B-V} and the calibration of Soderblom et al. (\cite{sod93}), would not
eliminate the discrepancy although, for some of the stars, it would
result in differences of $\sim~200$~K with respect to the
temperatures from {\it b-y} and the calibrations of Alonso et al. (1996,
1999). For example, Randich et al. (\cite{R02}) estimated \teff\ of
6061, 6016, and 6110~K for E7, E45, and T2105, still 250-300~K 
below our spectroscopic temperatures. 

Thus, problems related to the
{\it (b-y)} photometric scale are unlikely to be responsible for the
discrepancy. We also point out that the spectroscopic temperatures of
the main sequence stars cannot be strongly in error, given the excellent
agreement in the [Fe/H] between giants and dwarfs; were the \teff\ from 
photometry truly too low by 250 K, the dwarfs would
have subsolar [Fe/H] abundances ([Fe/H] $\sim -0.2$), in conflict with 
the giants and with the metallicity estimated from photometry.

\noindent {\it ii)} One could argue that the model atmospheres could be 
partly responsible for the discrepancy; in order to check this
possibility, we have analyzed the Sun and E45 with MOOG (Sneden
\cite{sne}) and the Kurucz (\cite{kuru}) model atmospheres. We find
a dependence of the difference in the derived abundances (ours$-$MOOG)
on both the line excitation potential and EWs, due to the fact that
Kurucz models are somewhat warmer and that damping is treated differently 
in the two codes; however, for the Sun we find an Fe abundance only 
0.03~dex higher ($\log~n$(Fe~{\sc i}) = 7.60), and for E45 [Fe/H] is 
exactly the same as found by us (0.15).  The \vt\ values agree within 
0.2 km/sec and \teff\ within $\sim$~50 K. The use of different model 
atmospheres and codes therefore does not introduce substantial 
differences in our results.

{\it iii)} We are therefore only left with the possibility that the
interstellar excess estimated from photometry is too low.  By adding
0.02 mag to the {\it E(b-y)} by Meibom, the two T$_{eff}$ are brought
almost into agreement. We conclude that {\it E(b-y)} = 0.091 or slightly 
larger for IC 4651. Note that such an increase in the reddening would 
not degrade the already acceptable agreement between spectroscopic and 
photometric temperatures for the giants, since their temperature scale is 
less sensitive than that of the dwarfs to {\it b-y} in this regime: with 
{\it E(b-y)}= 0.091, the spectroscopic and photometric temperatures of the 
six giants agree within 5 K.

Based on the above considerations and the more reliable and consistent 
abundances of the slowly-rotating giants and main sequence stars, we 
conclude that the Fe abundance of IC~4651 is [Fe/H]= +0.10$\pm$0.03.Using 
our higher value of {\it E(b-y)} would increase the photometric estimate of 
[Fe/H] by Meibom et al. (2002) by about 0.08 dex, still within the errors 
of both methods. The change in distance modulus is also within the errors 
and without consequence for our discussion.

\section{Other elements}

Individual element abundances for the stars in IC~4651, computed with the 
stellar parameters from the Fe analysis (Table~2), are given in Table~5. 
Table~6 lists mean abundances, element by element, for the Sun (both the 
canonical values from Anders \& Grevesse 1989 and our results), the 
main-sequence stars, and the giants in IC~4651. When computing the mean 
abundance for the giants we excluded E95, since for the discrepant elements 
(such as Na), this star shares the abundance pattern of the dwarfs rather 
than the giants. The abundances we derive from the UVES solar spectrum are 
all within 0.1 dex of the canonical solar values, and we use them as the 
reference for the cluster stars.

We have checked the sensitivity of our abundances to the assumed stellar
parameters by analyzing a giant (E60) and a dwarf star (E86) while varying 
\teff, log $g$, and \vt~separately. The results are given in Table~4. 

\begin{table}
\caption{Abundance changes ($\Delta$[X/H]) in two stars (E60 giant, E86 main sequence)
in response to variations of \teff, $\log g$, and \vt). }
\begin{tabular}{lllll}
\hline \hline
El. & T$_{eff} -$100 & T$_{eff} -$200 &$\log g -$0.3 & $\xi -$0.3\\ 
\hline      &        &  E60  &       &       \\     
  Al I  &  -0.06 & -0.13 &  0.01 &  0.05 \\
  Ca I  &  -0.10 & -0.21 &  0.02 &  0.12 \\
  Cr I  &  -0.15 & -0.32 & -0.02 &  0.22 \\
  Fe I  &  -0.05 & -0.08 & -0.03 &  0.13 \\
  Fe II &   0.13 &  0.27 & -0.16 &  0.11 \\
  Mg I  &  -0.03 & -0.04 & -0.01 &  0.02 \\
  Na I  &  -0.08 & -0.17 &  0.01 &  0.08 \\
  Ni I  &  -0.03 & -0.04 & -0.06 &  0.17 \\
  Si I  &   0.05 &  0.10 & -0.04 &  0.06 \\
  Ti I  &  -0.15 & -0.30 & -0.01 &  0.14 \\
  Ti II &   0.04 &  0.09 & -0.13 &  0.05 \\ \hline \hline
      &        & E86   &       &       \\
  Al I &  -0.03 & -0.07 & 0.01  &  0.01 \\
  Ca I &  -0.03 & -0.07 & 0.01  &  0.02 \\
  Fe I &  -0.05 & -0.12 & +0.01 &  0.05 \\
  Fe II &   0.01 &  0.02 & -0.09 &  0.08 \\  Mg I &  -0.03 & -0.07 &  0.01 &  
0.01 \\
  Na I &  -0.03 & -0.07 &  0.01 &  0.01 \\
  Ni I &   0.06 & -0.12 &  0.00 &  0.03 \\
  Si I &   -0.02 &  -0.06 & 0.03 &  0.03 \\
  Ti I &  -0.08 & -0.16 & 0.01 &  0.01 \\
  Ti II &   -0.01 &  -0.02 & -0.20 &  0.08 \\ 
\hline \hline

\end{tabular}
\end{table}

\begin{table*}
\caption{Element abundances for individual solar-type dwarfs and giants in 
IC~4651. Entries with ":" refer to uncertain measurements. For Cr, see text. } 
\begin{tabular}{lccccccccccccccc}
\hline
\hline
 Elem.  & E45   & E45 & E7 & E7 &  T2105 & T2105 &  E86 & E64 &  E25 &  E12 & 
E98 &  E60 &  E812 &  E95 &  E8  \\
\hline 
[Na I/H] & 0.02  & 0.06  & 0.14 & 0.11 & -0.10    & -0.03 & -0.18: & 0.10 & 0.08 
& 0.34 & 0.30 & 0.28 & 0.35 & 0.03 & 0.27 \\
\hline 
[Mg I/H] & 0.12  & 0.17  & 0.18 & 0.19 &  0.24   & 0.25  &   0.25 & 0.22 & 0.24 
&  0.17&  0.22&  0.18 & 0.21 & 0.31:& 0.18 \\
\hline
[Al I/H] & 0.00  & 0.01 & 0.08 & 0.08 &  0.03    & 0.05  &  0.11 &   -0.01& 0.08 
&   0.18 & 0.24 & 0.13 & 0.19 & 0.10 & 0.19 \\
\hline
[Si I/H] &   0.17&  0.18&  0.23&  0.19&  0.06   &  0.07 &    0.27 &   0.24 & 
0.25 &   0.15&  0.21 & 0.20 & 0.19 & 0.23 & 0.14 \\
\hline
[Ca I/H] &   0.15&  0.15&  0.15&  0.15 &  0.12  &  0.18  &   0.19 &   0.14 & 
0.21 &  0.12  & 0.11 &0.04 & 0.09 & 0.17 & 0.09 \\
\hline
[Sc II/H] &  -0.04&  -0.02& -0.06& -0.09& -0.26  & -0.21  &  -0.19 &   0.04 & 
0.05 &  0.03  & 0.19 & 0.15 & 0.07& -0.10& 0.02 \\
\hline[Ti I/H] &   0.14&  0.16 & 0.19&  0.14&   0.25 &   0.27 &   0.18 &   0.35: 
& 0.33:&   0.17 & 0.25&  0.24&  0.20&  0.10&  0.23 \\
\hline
[Ti II/H] &  -0.07& -0.06& 0.01 & 0.01 & -0.08   & -0.06  &  0.13  &  0.16 & 
0.26: &   0.04& 0.23 & 0.19 & 0.16 & 0.11& 0.10 \\
\hline
[Ni I/H] &   0.09&   0.15& 0.13&  0.16&   0.03  &  0.07  &  0.09 &   0.16 & 0.20  
&  0.14 & 0.28 & 0.24&  0.15&  0.05&  0.20 \\
\hline
[Cr I/H] &   0.18 &   0.20& 0.26&  0.20&   0.18  &  0.29  &  0.47:&   0.37: & 
0.71: & 0.08 & 0.12 & 0.11 & 0.01 & -0.17& 0.08 \\
\hline \hline 
\end{tabular}
\end{table*}
	
\begin{table*}
\caption{Average element abundances in the sun and in IC~4651, separately for 
the dwarfs and giants; numbers in parenthesis gives the average number of lines 
used in the mean. [Ti II/Fe] is relative to the solar Fe II abundance. } 
\begin{tabular}{lccccccl}
\hline \hline
\\ 
Elem. & \multicolumn{2}{c}{Sun} & [X/H]$_{\rm MS}$ & [X/H]$_{\rm giants}$ & 
[X/H]$_{\rm Cluster}$ & [X/Fe]$_{\rm MS}$ & [X/Fe]$_{\rm giants}$   \\
& A\&G & Us & & &  &  &    \\
\hline
Fe~{\sc i} (100) & 7.52 & 7.57 &  0.11  &  0.10 &  0.11  &        &       \\
Fe~{\sc ii} (7)  & 7.52 & 7.59 &  0.02  & -0.04 &   /    &        &       \\
Na~{\sc i}  (2)  & 6.33 & 6.28 &  0.02  &  0.29 &   /    & -0.09  &  0.19 \\
Mg~{\sc i}  (2)  & 7.58 & 7.53 &  0.24  &  0.19 &  0.21  &  0.13  &  0.09 \\
Al~{\sc i}  (2)  & 6.47 & 6.45 &  0.04  &  0.17 &  0.11: & -0.07  &  0.07 \\
Si~{\sc i}  (5)  & 7.55 & 7.45 &  0.18  &  0.18 &  0.18  &  0.07  &  0.08 \\
Ca~{\sc i} (17)  & 6.36 & 6.38 &  0.15  &  0.10 &  0.13  &  0.04  &  0.00 \\
Sc~{\sc ii} (2)  & 3.10 & 3.14 & -0.09  &  0.09 &  0.0:   & -0.11  &  0.11 \\
Ti~{\sc i} (21)  & 4.99 & 4.99 &  0.19  &  0.22 &  0.20  &  0.08  &  0.12 \\
Ti~{\sc ii} (5)  & 4.99 & 4.88 &  0.02  &  0.14 &  0.08  &  0.00  &  0.18 \\
Ni~{\sc i} (15)  & 6.25 & 6.23 &  0.12  &  0.20 &  0.16  &  0.01  &  0.10 \\
Cr~{\sc i}  (2)  & 5.67 & 5.72 &  0.22  &  0.08 &  0.15  &  0.11  & -0.02 \\
\hline \hline 
\end{tabular}
\end{table*}

\subsection{Comments on Abundances }

\noindent {\it Titanium and Scandium:} The Ti~{\sc ii} and Sc{\sc ii} abundances 
do not follow that of Fe~{\sc ii}; Fe~{\sc ii} appears to be underabundant in 
giants with respect to dwarfs (see Table 2), but the opposite is true for 
Ti~{\sc ii} and Sc{\sc ii}. While only two lines are available for Sc{\sc ii}, 
the number of lines of are comparable for Ti~{\sc ii} and Fe~{\sc ii}, so there 
seems to be no simple explanation for this difference. If some overionization 
was present, we would indeed expect the ionized species to be more abundant in 
giants than in dwarfs, but this should be true for Fe~{\sc ii} as well. A 
possible cause might be the strong dependence of Fe~{\sc ii} on \teff\ and $\xi$ 
in the giants, as emerging from the error analysis of Table 4. Since Ti~{\sc ii} 
is much less sensitive to the atmospheric parameters other than gravity, some 
small mismatch in these may be enough to cause the different behaviour. 

\noindent {\it Magnesium:} For Mg we only used the two lines at 6318.72 and 
6319.24 \AA. 

\noindent {\it Sodium:} Na shows the largest difference between giants and 
dwarfs. In order to convince ourselves of the reality of this discrepancy, we 
have first checked how other effects, such as non-LTE, may affect our abundance 
computation. Gratton et al. (1999) published NLTE calculations for dwarfs and 
giants and showed that a small correction, of order 0.07 dex, is required for 
giants with log $g\sim3$, such as those analysed here. A similar correction 
applies to the dwarfs, however, so NLTE is not the cause of the discrepancy. 
Our analysis of the effects of changing the stellar parameters shows that the 
only way to reduce this difference would be to lower the temperature scale for 
the giants by about 200 K. While temperatures 50 K cooler would help to bring 
the Fe ionization in balance in giants, values much larger would cause produce   
large discrepancies between dwarfs and giants in other elements, notably Ca, Ti, 
and Fe. We consider this possibility extremely unlikely and tentatively conclude 
that the difference in Na between giants and dwarfs is real, at the level of   
$\sim$0.2 dex. We note that the only subgiant in the sample, E95, has a Na 
abundance compatible with that of the dwarfs. 

\noindent {\it Aluminium:} Al is stronger in giants than in dwarfs; the 
discrepancy is marginally within the errors, however. NLTE effects must also be 
taken into account for Al: According to Baum\"uller and \& Gehren (1997), NLTE 
effects should increase the Na abundance in stars near solar metallicity when 
going from the main sequence to log $g$= 3; they would therefore make the 
discrepancy even larger. 

\noindent {\it Chromium:} In three stars (all dwarfs) we could measure only the 
line at 6362.88 \AA. In the similar star T2105, this line gives an abundance 
about 0.2 dex higher than the other line used, at 6330.10 \AA. The Cr I 
abundances given for these stars may therefore be about 0.2 dex higher than 
the real values and have not been used in the final computation. 

\noindent {\it Lithium:} Li deserves particular attention, since the abundance 
is computed from only one line, and Li is expected to vary along the CMD. [Li/H] 
has been determined from the 6708~\AA\ resonance doublet using MOOG; the 
equivalent width for this line was measured both automatically and manually for 
all the stars. In some cases, differences as large as 15 m{\AA} between the two 
measurements were found; in those cases, the manual measurements were repeated 
and found to be more reliable than the automatic ones; this is explained by the 
fact that the Li doublet is broader than normal single lines, so our version of 
DAOSPEC likely considered the Li line as a blend and underestimated its 
equivalent width. 

S/N ratio estimates indicate upper limits to the measured equivalent widths as 
small as 1-2 m{\AA} in sharp-lined stars, but, because most of our stars rotate 
quite fast, we adopted conservative upper limits of 5~m\AA.  For rapid rotators 
the measured equivalent width includes the contribution of the Fe 6707.46 line; 
for these stars, the measured EWs were corrected using the analytical 
approximation by Soderblom et al. (\cite{sod93}). {\it B-V} colors were 
corrected for reddening using our revised estimate of {\it E(b-y)}= 0.091, 
corresponding to {\it E(B-V)}= 0.13.  

The Li abundance is very sensitive to the adopted \teff; we have computed it 
using both the spectroscopic \teff\ from the Fe analysis and the photometric 
\teff\ using the Alonso scale and {\it E(b-y)}= 0.091 (see above). For most 
stars the difference is negligible, and in no case will it influence our 
conclusions. We estimate the uncertainty in \teff\ to be about 200 K for dwarfs 
and TO stars and half that amount for the giants, corresponding to typical 
uncertainty of 0.2 dex in [Li/H] for all the stars. However, most of this is due 
to the uncertainty in  {\it E(b-y)}, and a change in {\it E(b-y)} would produce 
an almost constant shift of the Li abundances without affecting most of our 
conclusions, which are based on the pattern of changes in Li along the CMD.

Table 7 reports the Li equivalent widths and abundances for all the stars. 
Asterisks denote stars with unblended Li lines; for the other stars, the 
equivalent widths include the Fe I blend.

\begin{table}[h]
\caption{Lithium abundances for the stars in IC4651, computed with both 
spectroscopic (S) and photometric (P) values of \teff, assuming a reddening of 
{\it E(b-y)}= 0.091; we use the latter in our analysis. Asterisks indicate that 
W(Li) (in m\AA) refers to the unblended Li line. }
\vspace{-0.5cm}
$$
\begin{array}{lccrrr}
\hline \hline 
Star  & T(S)  & T(P) & W(Li) & \log N(Li)(S) & \log N(Li)(P) \\
\hline
E12   & 5000 & 5020  &   <5   & < 0.35 & < 0.38 \\
E8    & 4900 & 4950  &   <5   & < 0.09 & < 0.23 \\
E98   & 4900 & 4950  &   16 * &   0.72 &   0.82 \\
E60   & 4900 & 4930  &    9 * &   0.42 &   0.53 \\
E812  & 5000 & 4910  &   <5   & < 0.38 & < 0.21 \\
E95   & 5800 & 5670  &   35 * &   2.19 &   2.07 \\
E3    & 6550 & 6550  &    9   &   1.64 &   1.64 \\
E56   & 6500 & 6520  &   <10  & < 2.15 & < 2.16 \\
E19   & 6500 & 6280  &   <5   & < 1.83 & < 1.66 \\
E99   & 6800 & 6830  &   <10  & < 2.36 & < 2.38 \\
E25   & 6900 & 6680  &   75   &   3.44 &   3.29 \\
E5    & 7100 & 6930  &   76   &   3.59 &   3.47 \\
T1228 & 6800 & 6770  &   58   &   3.20 &   3.18 \\
E14   & 6800 & 6860  &   <5   & < 2.03 & < 2.07 \\
E34   & 6850 & 6640  &   <5   & < 2.07 & < 1.92 \\
E15   & 6850 & 6540  &   <6   & < 2.15 & < 1.93 \\
E79   & 6650 & 6620  &   <5   & < 1.93 & < 1.91 \\
E64   & 6650 & 6550  &   10 * &   2.24 &   2.17 \\
E86   & 6600 & 6459  &   10 * &   2.20 &   2.13 \\
T2105 & 6350 & 6138  &   67 * &   3.02 &   2.84 \\
E7    & 6300 & 6308  &   53 * &   2.83 &   2.84 \\
E45   & 6350 & 6103  &   47 * &   2.80 &   2.59 \\
\hline \hline
\end{array}
$$
\end{table}

\section{Discussion}

\subsection{Chemical composition of IC 4651}

Our mean abundances for Fe and several other elements in IC4651 are all reported 
in Table~6. For most elements, accurate mean abundances for the whole cluster 
can be derived, given the very good agreement between the results for individual 
stars.  In the following we will use the results for disk stars by Edvardsson et 
al. (1993) and Chen et al. (2000) as our reference in the comparisons.

\subsubsection{$\alpha$ elements: Mg, Si, Ti and Ca }

[Mg/Fe] is the same for giants and dwarfs and slightly above Solar ([Mg/Fe]= 
0.11). Considering that the cluster is metal rich, this fits the general trend 
of [Mg/Fe] in disk stars extremely well. In their discussion, Chen et al. (2000) 
conclude that the mechanisms of Mg production are not well understood, as some 
Mg seems to be produced much after SNe II cease to dominate the enrichment.

[Si/Fe] and [Ti/Fe] are also very well defined and slightly positive (0.07 -
0.08), like Mg. Together, the slight overabundance of all these elements is 
highly significant and fits the trend in field stars of comparable metallicity.
In contrast, [Ca/Fe] is not significantly different from zero, so Ca seems not 
to share the overabundance of the other $\alpha$-elements. 

Together, our results strongly indicate that none of the $\alpha$-elements is 
underabundant at this metallicity, so that the [$\alpha$/Fe] relation flattens 
out, or even increases, at high metallicity.

\subsubsection{Sodium and aluminium }

The behaviour of Na and Al is unfortunately less clean, mostly because of the 
abundance differences between giants and dwarfs. Our dwarf data would indicate 
that Al is slightly underabundant, at odds with what is found among field stars 
(Chen et al. 2000), while the low value of [Na/Fe] would completely agree with 
the field data, which are flat and scattered above the Solar [Fe/H]. We note 
also that low Na and Al are observed for the young stars in the Chen sample, 
which would be in better agreement with our data for the relatively young IC 
4651. Given the discrepancy between giants and dwarfs, we can only conclude that 
the ratio of these elements to Fe is solar or slightly below for IC 4651.

For Na, we note that the dwarfs and the low-luminosity giant of the sample (E95) 
share the same low Na value, while all the other giants have much higher Na. We 
believe that this [Na/Fe] enhancement in the giants with respect to the dwarfs 
is real. A possible Na overabundance of the order of 0.2 dex has been identified 
also by Tautvaisiene et al. (2000) among the clump giants of M67, very similar
to what we find in IC 4651 (0.28 dex). On the other hand, the same authors found 
no O deficiency nor any Al overabundance in their stars, showing that if 
the Na overabundance is due to internal nucleosynthesis plus mixing, then this 
process is not accompanied by the MgAl cycle. This would agree with other 
studies, which have shown that the MgAl cycle dominates at lower metallicity.

Na overabundances have been observed in A-F supergiants (Takeda \& Takeda-Hiday 
1994 and references therein). They are explained simply by a first-dredge up of 
$^{23}$Na, synthesized by the Ne-Na mode of H-burning while the star was on the 
main sequence (El Eid \& Champagne 1995). Theory predicts that the post 
dredge-up enhancement of Na should increase with stellar mass so much less 
$^{23}$Na should be synthesized inside the present giant stars in IC4651 than in 
these A-F supergiants: Our 3.0M$_{\odot}$ and 2.2M$_{\odot}$ models (see \S 
6.2.2) predict a Na enhancement of 0.3 dex and 0.12 dex, respectively, with 
respect to the initial main-sequence value. 

Given the substantial uncertainty of the nuclear reactions involved in the 
$^{23}$Na production (Arnould et al. 1999) as well as the errors of our Na 
abundance determinations (0.15 dex for a \teff\ error of 100 K in the giants), 
we see no serious conflict between first dredge-up predictions and the observed 
Na abundances in the giant stars. We note, however, that if the reported 
difference in M67 is confirmed, this interpretation might be more problematic, 
given the substantially lower TO mass of the older M67.

\subsubsection{The Fe group: Cr and Ni }

With [Ni/Fe]= [Cr/Fe]= 0.05, both elements share the overabundance of Fe and the 
general trends found in field stars.

\subsection{Lithium }

\subsubsection{Li abundances along the cluster sequence }

\begin{figure*}[th]
\begin{center}
\includegraphics[width=16.0cm,height=12.0cm]{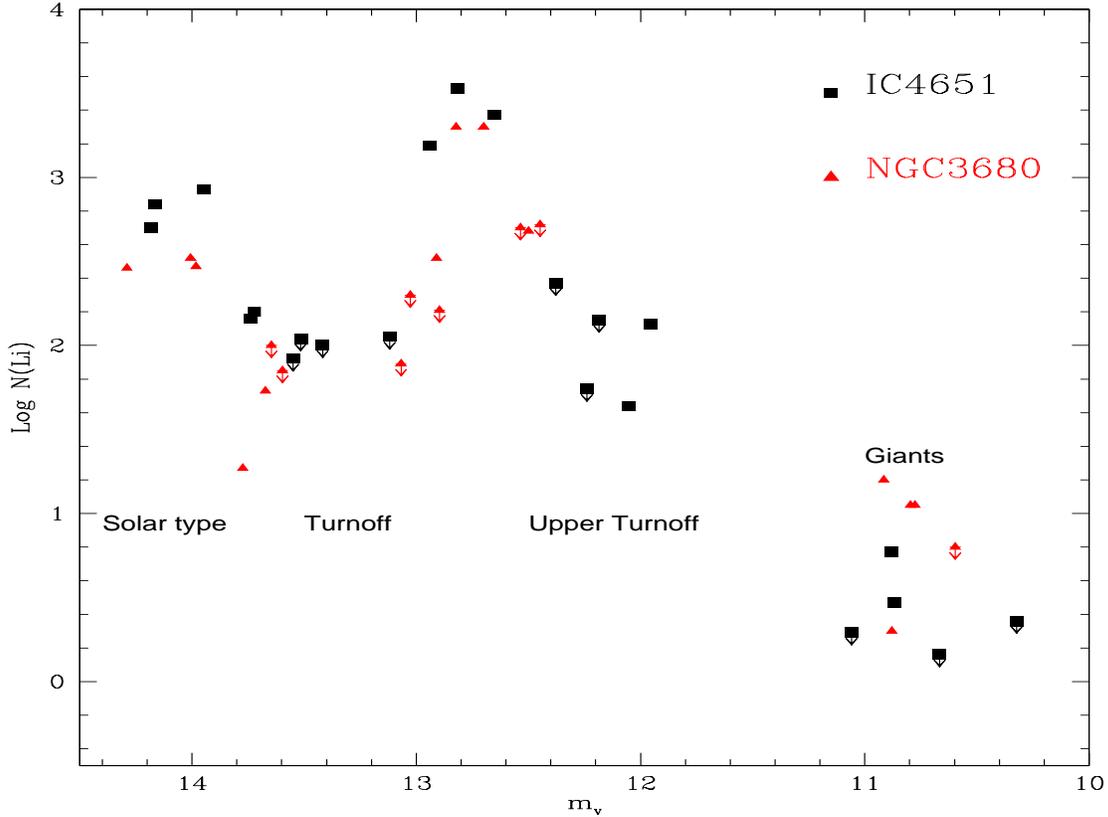}
\end{center}
\caption{Li abundance vs. $m_V$ for stars in IC 4651 (squares) and NGC 
3680 (triangles). 
}
\end{figure*}

In a detailed study of the intermediate-age open cluster NGC 3680, Pasquini et 
al. (2001) found, in addition to the standard features (i.e. the Li dip and the 
plateau for solar-type stars), that: {\it a):} Li reaches {\it young Pop. I} 
values on the bright side of the dip; {\it b):} Li drops dramatically in even 
brighter stars, as soon as they become slightly evolved; and {\it c):} Giants 
very close together in the CMD show substantial scatter in Li abundance, with 
the clump giants having a measurable amount of Li. 

Given its similar age to NGC3680 (within 0.1-0.2 Gyr, Meibom et al. 2002), IC 
4651 is a perfect object with which to test whether the results for NGC 3680 may 
apply to other clusters, and thus potentially to entire disk population of age  
$\sim$ 1.5-1.7 Gyr.

In Fig. 3 we plot the Li abundance vs. m$_{\rm V}$ magnitude for both clusters 
(which are also at the same distance). The Li data for IC 4651 are from Table 7 
(based on the photometric \teff), with an estimated uncertainty of 0.2 dex; the 
NGC 3680 data are from Pasquini et al. (2001). For comparison between the 
clusters this is preferable to the classical Li vs. \teff\ diagram because of 
the degeneracy of the latter in the critical turnoff region (Pasquini et al. 
2001). 

The different sections of the CMD are also indicated in Fig. 3. Moving from left 
to right, the diagram thus shows what happens in stars with increasing initial 
mass.

The Li abundance is an extremely tight function of $m_V$, confirming that 
the pattern observed in NGC 3680 is indeed common to both clusters; viz.:

\begin{itemize}
\item Stars fainter than the Li dip (the 3 solar-type stars E7, E45 and T2105 
analysed by Randich et al. 2000) have roughly the same Li abundance, slightly 
below the younger Hyades stars at the same temperature. Note that our values 
differ from those by Randich et al. (2000) because of the substantially higher
\teff\ values adopted in the present work. The general discussion of solar-type 
stars in clusters in Pasquini et al. (2000) and Randich et al. (2000, 2002) will 
not be repeated here. 

\item The Li dip is extremely very well defined, although for some of our stars   
we can only give upper limits to the Li abundance.

\item Immediately above the dip, the slightly more luminous stars (E5, E25, and 
T1228) have the meteoritic Li abundance (e.g. the initial Li content of the 
solar system, [Li/H]$\sim$3.3) or even higher. This is near the bluest point on 
the TO, before the redward ``hook" in the isochrones. Given the uncertainties in 
\teff, we do not consider the apparent super-meteoritic value to be significant. 
It is clear, however, that these stars have not suffered any significant surface 
depletion of Li during their main-sequence lifetime.

\item In dramatic contrast, the Li abundance drops precipitously by a factor 
of 10 or more in the stars that are just 0.25 mag brighter (E99 and the slightly 
brighter E3, E19 and E56).

\item Finally, the giants exhibit an extremely interesting behaviour: Out of 
five stars, three have only upper limits while Li is clearly detected in the two 
others, which are located in the clump region of the CMD. 
\end{itemize}

In the following subsection we return to these patterns, which appear to be 
signatures of non-standard mixing processes in the stars.

\subsubsection{Comparison with the predictions from rotating stellar models}
The study of Li in open clusters is a very powerful tool to investigate the 
dynamical phenomena in the stellar interiors 
(e.g., Pasquini 2000, Deliyannis et al. 2000, Randich et al. 2000). In Talon 
\& Charbonnel (1998, hereafter TC98) and Palacios et al. (2003, hereafter 
PTCF03), measurements of Li abundances and rotational velocities in young open 
clusters ($\alpha$ Per, Pleiades, Ursa Major, Coma Berenices, and the Hyades) 
were used to test the predictions of rotating stellar models including the most 
recent theoretical developments concerning meridional circulation, turbulence by 
shear instability, and atomic diffusion (see the above papers for a detailed 
description of the treatment of rotational mixing). 

TC98, in particular, showed that the hot side of the Li dip in these young 
clusters is well described by these models, which also successfully reproduce 
the He, C, and N anomalies in O-type and early B-type stars (Talon et al. 1997) 
and in O supergiants, as well as the B depletion in main-sequence B-type stars
(see Maeder \& Meynet 2000 for references). Charbonnel \& Talon (1999, hereafter 
CT99) and PTCF03 extended the comparison between model predictions and 
observations of Li to evolved stars originating from the hot side of the dip. 
At that time, however, the necessary data along the subgiant branch were 
available only for field stars (Wallerstein et al. 1994, L\`ebre et al. 1999, do 
Nascimento et al. 1999, Randich et al. 1999), while the data for open clusters 
concerned only bright giants (Gilroy 1989). The comparison between the models 
and the data was very satisfactory.

Our present analysis, together with the previous results for NGC 3680, fills 
several gaps in the observational picture and provides tighter constraints on 
the models. On the one hand, we now have lithium data for main-sequence 
stars in open clusters with an age intermediate between the very well-studied 
young population (younger than the Hyades) and the old cluster M67. On the other 
hand, our abundance determinations trace the behaviour of lithium along the 
subgiant branch more precisely. This allows us to test the rotating models as 
regards the age-dependence of the Li dip and the Li post-main-sequence evolution 
simultaneously. 

In the following we focus our attention on the stars on the blue side of the Li 
dip, ignoring the main-sequence stars cooler than 6500 K. For these ``solar-type 
stars" indeed it has already been shown (Pasquini et al. 1994, Randich et al. 2000, 
2002) that current models do not match the Li and Be observations.
TC98 discussed this problem in terms of an additional mechanism,  
possibly related to internal gravity waves, which transports 
angular momentum efficiently in these lower-mass stars while it is inefficient 
on the blue side of the dip (see also CT99 and TC03 for more details). 
%
New models for low-mass stars of the red side of the dip, 
including angular momentum transport by gravity waves in addition 
to the hydrodynamical processes induced by rotation, 
are not yet available. 

Here, we compare the predictions of the CT99 and PTCF03 models 
for stars on the blue side of the dip with the Li data 
in IC 4651 in the classical way, i.e., as functions of \teff\, separately for 
the main-sequence and turnoff stars (Fig. 4) and for the evolved stars (Fig. 5). 
The small differences in the description of the rotation-induced mixing in the 
two sets of models (see PTCF03 for a detailed discussion) enable us to combine 
their predictions and compare them to the present data. The initial rotation 
velocity used in both studies was very similar (100 or 110 km s$^{-1}$), and 
the stars were assumed to undergo magnetic braking leading to the observed 
rotation velocities at the age of the Hyades. 

\noindent {\bf \underline {Main sequence and turnoff stars}}

Fig. 4 compares the predictions of the models for an age of 1.5 Gyr to our 
observational data for the main-sequence and turnoff stars in IC 4651. Our 
photometric values of \teff\ and the corresponding log N(Li) are shown 
(black points: actual determinations; black triangles: upper limits). Open symbols 
show the model predictions at 1.5 Gyr. The open circles correspond to models 
with an initial rotation velocity of 110 km s$^{-1}$. For the 1.5M$_{\odot}$ 
model, this circle is for an initial $\rm V_{rot}$ of 100 km s$^{-1}$, and it is 
connected to the predictions for an initial velocity of 50 km s$^{-1}$ (open 
triangle) and 150 km s$^{-1}$ (open square). These points provide an estimate 
of the expected dispersion at an age of 1.5 Gyr for stars inside the Li dip 
which have undergone relatively strong braking early on the main sequence; it 
should be lower for more massive stars with thinner convective envelopes, for 
which the braking is less efficient. As can be seen, the models that already 
explained the shape of the hot side of the Li dip in the young open clusters 
also successfully account for the Li behaviour at later ages.

We note that the rotating models have been computed without 
adhoc convective core overshooting.
However, rotational mixing naturally extends the convective core compared to the 
classical case, and has a similar effect as overshooting on the main-sequence 
lifetime and the turnoff shape in the HRD (this also explains why models for a 
given initial mass, but different initial rotational velocities show a range in 
\teff\ at a given age, as seen in Fig. 4 
for the 1.5M$_{\odot}$ model). More extended grids of rotating 
models and corresponding isochrones now need to be computed and compared to 
colour-magnitude diagrams of open clusters with various ages.

\begin{figure}[h]
\begin{center}
\includegraphics[width=8.0cm,height=8.0cm]{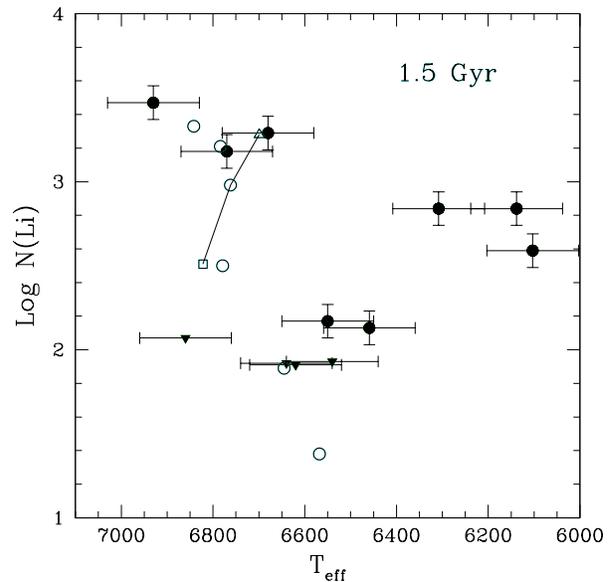}
\end{center}
\caption{Li abundance in main-sequence and turnoff stars fainter than $V 
\sim$12.6 in IC 4651 (black points: actual determinations; black triangles: upper 
limits) vs. \teff\ from photometry. Open circles show model predictions from 
CT99 and PTCF03 for an initial $\rm V_{rot}$ of 110 km s$^{-1}$ and different masses. 
For 1.5M$_{\odot}$, additional models for an initial $\rm V_{rot}$ of 50 and 150 km 
s$^{-1}$ are also shown (open triangle and square, respectively).}
\end{figure}

\begin{figure}[h]
\begin{center}
\includegraphics[width=8.0cm,height=8.0cm]{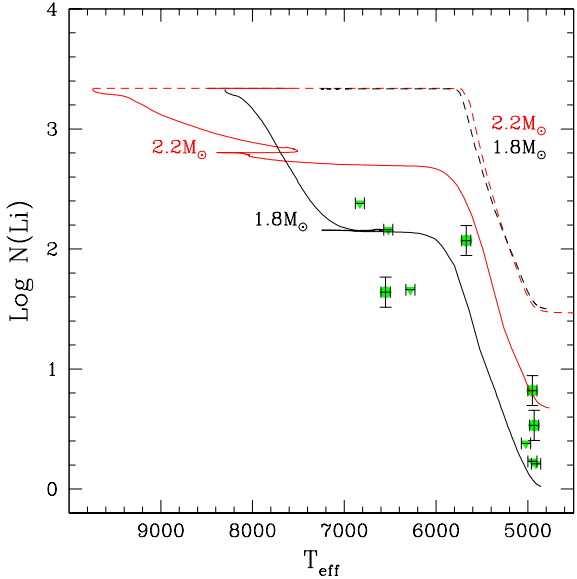}
\end{center}
\caption{Li abundances in post-turnoff stars brighter than $V \sim$12.6 and 
giants in IC 4651 (photometric \teff\ values). The lines show the evolution of 
surface Li in standard models (dashed) and in two rotating models computed with 
an initial $\rm V_{rot}$ of 110 km s$^{-1}$ (PTCF03; full lines).}
\end{figure}

\noindent {\bf \underline {Subgiant stars}}

We now turn to our sample of post-turnoff stars in IC 4651. Fig. 5 compares 
their Li abundances as a function of \teff\ with stellar models from PTCF03. The 
dashed lines show the evolution of the surface Li abundance in standard (i.e. 
non-rotating) models for 1.8 M$_{\odot}$ and 2.2 M$_{\odot}$. In both cases, 
surface Li depletion only begins inside the Herzsprung gap at $\sim$5800 K and 
is simply due to dilution as the convective zone deepens. The observed drop in 
lithium at much higher \teff\ cannot be explained by just this classical effect. 
Exactly the same behaviour is observed in NGC 3680 stars and in other clusters 
with turnoff masses above $\sim$1.6 M$_{\odot}$ (Gilroy 1989; Pasquini et al. 
2001) and in field stars (see PTCF03 and references therein).  

This points to a mechanism which is characteristic of stars originating from the 
hot side of the Li dip and is already efficient on the main sequence, even if 
its signature appears only as the stars start to cross the Hertzsprung gap 
(Vauclair 1991; Charbonnel \& Vauclair 1992). By enlarging the Li-free regions 
inside these objects relative to the classical case, rotation-induced mixing 
actually prepares the stars for the Li abundance variations seen in later 
evolutionary phases. The predicted behaviour of their surface Li is shown in 
Fig. 5 (fulllines). 

As can be seen, the rotating model for 1.8 M$_{\odot}$ (the approximate turnoff 
mass of IC 4651) exhibits earlier surface Li depletion than in the classical 
case (i.e., at higher \teff). This model, computed with an initial rotation 
velocity of 110 km s$^{-1}$ as described previously, also explains the data for 
field subgiant stars, as discussed in CT99 and PTCF03 (see Fig. 10 of PTCF03).
The present comparison shows that it also nicely matches the data for the 
subgiant stars in IC 4651. As already discussed for our main-sequence stars, 
one expects a dispersion in initial $\rm V_{rot}$ which will account for the 
observed dispersion in the sample. 

\noindent {\bf \underline {Giant stars}}

We end our discussion with the giant stars in IC 4651, which are also included 
in Fig. 5 
(data points at 4900-5000K). 
As indicated by isochrone fits to the cluster (see e.g Meibom et al. 
2002), these stars have masses 
higher than 1.8 M$_{\odot}$ 
(the exact mass range depending on the physics included in the various sets of models). 
As can be seen, standard models for these objects lead to post-dilution Li 
abundances that are much higher than observed. 
The rotating models, however, predict a total Li 
depletion after dredge-up that is much larger than in the standard case, and 
again fit the observations very satisfactorily. The conclusions by CT99 and 
PTCF03, that the rotating models can explain the mean Li depletion and abundance 
dispersion in evolved stars above $\sim$1.4 M$_{\odot}$ in the field and in open 
cluster are thus confirmed.


Cluster giants were discussed extensively by Pasquini et al. (2001), who
speculated that clump giants could show more Li than stars on the first ascent
of the RGB. Our observations of IC 4651 could possibly
support this hypothesis. Indeed also in this cluster 
we observe only upper limits for 
the stars that are found away from the clump,  
which are lower than our typical detections in the stars in the clump region.
Again, rotating models could provide a plausible explanation:
in a given cluster, the clump giants had slightly higher initial masses than the RGB stars, 
and both theory and observations imply that they have undergone more modest 
braking while on the main sequence; thus, retaining more Li. 
This can be seen in Fig. 5, where the Li abundance after dredge-up in the 2.2 M$_{\odot}$ 
rotating model is higher than in the 1.8 M$_{\odot}$ model and fits the clump data perfectly.
However, in a given cluster the difference in masses between RGB and
clump stars can be at most of 0.1 M$_{\odot}$, which does not seem enough to justify
the difference. 

The real point is that in the considered clusters it is very difficult to
firmly establish which really are the clump stars. 
Antony-Twarog et  al. (2004) conclude for instance that the high Li stars 
in NGC3680 could well be RGB objects rather than clump stars. 
In view of these considerations we consider the previous hypothesis as rather unlikely.

Another possibility is that some Li is produced after the end of the first 
dredge-up. The RGB bump is the most attractive phase where this might take 
place, as shown by Charbonnel \& Balachandran (2000). At that time, the 
outward-moving hydrogen burning shell burns through the mean molecular weight 
discontinuity created by the first dredge-up. 
An extra-mixing process can then easily connect the $^3$He-rich envelope 
material to the outer regions of the hydrogen burning shell, enabling Li 
production to occur. Very high lithium abundances can then be reached, as 
confirmed by the existence of the so-called ``lithium-rich giants". This phase 
is extremely short-lived and is followed by a phase where deeper mixing destroys 
the freshly synthesized Li and lowers the carbon isotopic ratio (e.g., 
Charbonnel et al. 1998). Palacios et al. (2001) discussed this extra-mixing 
process in the framework of rotation-induced mixing and called this episode the 
``lithium flash". 

In the present context, the fact that the giant stars in NGC 3680 and IC 4651 
have higher Li abundances than their counterparts in NGC 752 (see Pasquini et 
al. 2001) may be significant. As discussed by Anthony-Twarog \& Twarog (2004), 
the giants in NGC 3680, and possibly also in IC 4651, could be predominantly 
first-ascent giants in the bump phase, while the lithium-poor giants in NGC 752 
would be clump stars. It is worth noting, however, that these objects are just 
in the critical mass range where He ignition passes from a degenerate-core flash
to a quiescent process. The He flash occurs in less massive stars (found in 
older clusters like NGC 3680 and IC 4651), which will indeed go through the RGB 
bump and undergo some Li enrichment. The slightly more massive stars in younger
clusters like NGC 752 will ignite He before going through the bump, and thus 
without undergoing the Li flash. An analysis of the $\mathrm{C^{12}/C^{13}}$ 
ratio and Be abundance in the giants of these three clusters will allow to 
test if this explanation is correct.

\section{Conclusions} 
Our superb high resolution, high S/N ratio spectra from the ESO VLT/UVES of 23 
stars in the intermediate-age open cluster IC 4651 have enabled us to perform an 
accurate chemical analysis of stars in critical phases along the CMD of this 
cluster. Abundances of the Fe-peak and $\alpha$ elements and Li have been 
derived and interpreted with the help of state-of-the-art evolutionary models 
incorporating rotational mixing. 

We find that the reddening towards the cluster was previously slightly 
underestimated and we derive a new value of {\it E(b-y)}= 0.091. We confirm that 
the Fe abundance of IC 4651 is supersolar ([Fe/H]= 0.10]), and that the cluster 
shows a small $\alpha$-element enhancement. For most elements, the abundances of 
main-sequence and giant stars are in excellent agreement, but Na is overabundant 
in the giant stars with respect to the dwarfs. We interpret this as evidence for 
dredge-up of Ne processed material. 

We discuss Li extensively: The trends of Li abundance vs. mass and stage of 
evolution follow very closely the dramatic variations observed in the similar 
cluster NGC 3680 (Pasquini et al. 2001), confirming that our results for these 
clusters are typical of disk stars at an age of 1.6-1.8 Gyr. The observed 
behaviour of Li in stars more massive than those in the Li dip is very well 
reproduced by the most recent stellar evolution models which include rotational 
mixing.
Finally, the pattern of Li depletions in the 
giant stars is still quite puzzling, and its interpretation critically depends on our 
ability to discriminate between first-ascent RGB and clump stars. 

Further analysis of other elements in IC 4651 is in progress, but the present 
work has already made IC 4651 a member of the tiny club of open clusters with 
precise, detailed spectroscopic abundances. 

\begin{acknowledgements}
We thank S. Meibom for the opportunity to select our programme stars from his 
thesis in advance of publication, P. Stetson for similarly providing DAOSPEC 
before publication, and J. Andersen for fruitful discussions and comments on 
an earlier draft of the paper. BN thanks the Carlsberg Foundation, the Danish 
Natural Science Research Council, the Swedish Research Council, and the Nordic 
Academy for Advanced Study for financial support. We Finally thank the referee, 
B. Anthony-Twarog, for useful comments. 
\end{acknowledgements}

\end{document}